\documentclass[pre,twocolumn,aps,eqsecnum]{revtex4}
\usepackage{epsfig}

\begin{document}

\title{Ising Model of a Glass Transition}

\author{J.S. Langer}
\affiliation{Department of Physics, University of California, Santa Barbara, CA  93106-9530}

\date{\today}

\begin{abstract}

Numerical simulations by Tanaka and coworkers indicate that glass forming systems of moderately polydisperse hard-core particles, in both two and three dimensions, exhibit diverging correlation lengths.  These correlations are described by Ising-like critical exponents, and are associated with diverging, Vogel-Fulcher-Tamann, structural relaxation times.  Related simulations of thermalized hard disks indicate that the curves of pressure versus packing fraction for different polydispersities exhibit a sequence of transition points, starting with a liquid-hexatic transition for the monodisperse case, and crossing over with increasing polydispersity to glassy, Ising-like critical points.  I propose to explain these observations by assuming that glass-forming fluids contain twofold degenerate, locally ordered clusters of particles, similar to the two-state systems that have been invoked to explain other glassy phenomena.  This paper starts with a brief statistical derivation of the thermodynamics of thermalized, hard-core particles. It then discusses how a two-state, Ising-like model can be described within that framework in terms of a small number of statistically relevant, internal state variables.  The resulting theory agrees accurately with the simulation data.  I also propose a rationale for the observed relation between the Ising-like correlation lengths and the Vogel-Fulcher-Tamann formula.  

\end{abstract}

\maketitle

\section{Introduction}
\label{Intro}

Recent numerical simulations by Tanaka and coworkers  \cite{TANAKAetal-10,KAWASAKI-TANAKA-11,TANAKA-EPJ-12} indicate that glass forming systems of moderately polydisperse, thermalized, hard-core particles, in both two and three dimensions, exhibit diverging bond-orientational correlation lengths with Ising-like critical exponents. They also exhibit diverging relaxation times consistent with the Vogel-Fulcher-Tamann formula.  A separate  analysis by Mosayebi et al~\cite{MOSAYEBIetal-10}, for a three dimensional, bidisperse, Lennard-Jones model,  similarly produces an Ising-like correlation length, with an ordering mechanism that apparently is different from the bond-orientational one seen by Tanaka.  Taken as a whole, along with the pressure-density curves measured by Kawasaki and Tanaka  \cite{KAWASAKI-TANAKA-11} for hard disks at varying polydispersities, this emerging body of evidence raises the possibility that glass forming liquids exhibit some underlying universality.  My purpose in this paper is to explore one such possibility.  

It is not clear how the behaviors reported in  \cite{TANAKAetal-10,KAWASAKI-TANAKA-11,TANAKA-EPJ-12} can be consistent with existing glass theories.  They are reversible features of thermodynamically equilibrated states; therefore, it seems unlikely that they can be explained by facilitation models that have no nontrivial thermodynamic properties.\cite{FREDRICKSON-ANDERSEN-85}  
Mode coupling theory \cite{GOTZE91,GOTZE92}  does start with a well posed model of a thermally equilibrated molecular fluid, but its perturbation-theoretic nature prevents it from probing strongly collective phenomena at glass transitions. There is a large literature on topological constraint theories of molecular glasses, in which the constraints are imposed by chemical bonds.  (See \cite{MAURO-11} for a concise review.) The intuitively attractive concept of ``rigidity percolation'' that is central to the constraint theories may be related to the Ising-like correlations considered here. That conjecture deserves further study from a statistical thermodynamic point of view; but it is beyond the scope of the present discussion.  

The computational models used in  \cite{TANAKAetal-10,KAWASAKI-TANAKA-11} are systems of particles with short-ranged interactions and intrinsic fluidlike disorder.  Such models are physically very far from the spin-glass models with long ranged forces and quenched disorder that have been used, for example, as the basis for the random first-order transition theory (RFOT). \cite{RFOT-00-07,RFOT-89}  My own excitation-chain (XC) theory \cite{JSL-XCHAINS-06} originally was introduced as a way of avoiding the weaknesses I perceived in RFOT  \cite{JSL-DH-11}; but the diverging length scale predicted by the XC theory does not have Ising-like critical exponents. It seems to me, therefore, that Ising-like universality -- if confirmed --  must arise from some previously unexpected, fundamental feature of a large class of glass forming fluids.

Throughout this paper, I focus primarily on systems of thermalized particles interacting only via very short ranged, repulsive forces.  By ``thermalized,'' I mean that the particles have an average kinetic energy determined by their temperature or by the temperature of an inert fluid in which they are immersed.  In thermodynamic equilibrium, hard-core systems of this kind are characterized entirely by steric constraints and configurational entropy.  As remarked above, they are the antitheses of the models with infinitely long-ranged interactions that are used as starting points for some glass theories, e.g. RFOT.  They contain no stored elastic energy. Nor, for classical systems of this kind, can the entropy associated with kinetic degrees of freedom play any role in determining equilibrium structures; the kinetic energy simply factors out of the partition function.  Thus, hard-core simulations, and analogous experiments using hard-core colloidal particles, pose an especially  clean theoretical challenge. 

Section~\ref{Tanaka} of this paper contains a brief summary of the simulations reported in \cite{TANAKAetal-10,KAWASAKI-TANAKA-11,MOSAYEBIetal-10}  Then, in Sec.~\ref{Thermo}, I present a statistical derivation of the thermodynamics of thermalized, hard-core particles, which also serves to emphasize the role of properly chosen internal state variables. In the main part of this paper, I use this thermodynamic framework to develop an Ising-like theory of disordered, hard-core materials. The basic ingredient of this theory is a population of twofold degenerate, topologically oriented, clusters of particles, similar to the two-state systems that  have been invoked to explain other glassy phenomena.\cite{AHV-72,FL-11}   In Sec.~\ref{Twostate}, I present a rationale for this theory, and then, in Secs.~\ref{Ising} and \ref{Pphi}, show that it predicts both Ising-like, bond-orientational correlations, and a sequence of critical ordering transitions along the curves of pressures $p$ as functions of packing fractions $\phi$ for varying polydispersities. In Sec.~\ref{Rates}, I propose a rationale for Tanaka's observed relation between the correlation length and the Vogel-Fulcher-Tamann relaxation times.  

In Sec.~\ref{Twostate}, however, I also argue that Ising symmetry must cross over to full rotational symmetry when the correlations become sufficiently long-ranged near critical points.  As a result, the two-state theory must fail in the immediate vicinities of ordering transitions.  Nevertheless, this theory accounts remarkably well for the simulation results, and may therefore provide a unified point of view for understanding a wider range of glassy behaviors. I conclude in Sec.~\ref{Conclusions} with some remarks about that conjecture.

\section{Simulation Results}
\label{Tanaka}

In \cite{TANAKAetal-10}, Tanaka et al. report Brownian simulations of a variety of two and three dimensional systems, including (but not restricted to) polydisperse hard disks and spheres. They deduce bond-orientational correlation exponents by finite-size scaling analyses.  They also report measurements of relaxation times $\tau_{\alpha}$ as functions of packing fraction and degree of polydispersity.  As emphasized by Tanaka in \cite{TANAKA-EPJ-12}, the dramatic slowing down near the glass transitions prevents these simulations from coming sufficiently close to the critical points to confirm the conjectured limiting behaviors. They can observe growth of correlations by only about one decade at best. It seems to me, however, that the consistency between these results for a range of different models means that we may take them seriously, at least pending further study. The independent analysis by Mosayebi et al~\cite{MOSAYEBIetal-10}, for a three dimensional, bidisperse, Lennard-Jones system, adds weight to  this evidence.  In \cite{TANAKA-EPJ-12}, Tanaka reports seeing no bond-orientational order in glass forming, binary systems. In \cite{MOSAYEBIetal-10}, however, Ising-like correlations are observed by looking at distant, non-affine, linear responses to local perturbations, which could mean that some different kind of ordering mechanism is operative here.  If so, there is additional reason to look for the origin of this universality.

The Ising-like picture that emerges in both \cite{TANAKAetal-10} and \cite{MOSAYEBIetal-10} is that the  correlation lengths $\xi$ are proportional to $t^{- \nu}$ with $\nu = (2/d) - \alpha$ (hyperscaling), where $d$ is the spatial dimensionality, $\alpha$ is the specific-heat exponent (negligibly small for these purposes), and $t$ is a dimensionless measure of the distance from a critical point.  For thermally controlled systems, $t = (T -T_c)/T_c$, where $T_c$ is the critical temperature.  For hard-core particles,  the temperature $T$ is replaced by the inverse of the packing fraction $\phi$.  For both $d = 2 ~{\rm and}~ 3$, the structural relaxation time $\tau_{\alpha}$ is found to be consistent with a Vogel-Fulcher-Tamann (VFT) relation, $\log\,(\tau_{\alpha})\sim \xi^{d/2} \sim t^{-1}$. 

More evidence bearing on the phase transitions that occur in these systems is summarized in Fig.~2 of \cite{KAWASAKI-TANAKA-11}.  Here, Kawasaki and Tanaka have plotted the pressure $p$ as a function of $\phi$ for simulated hard disks at a sequence of increasing percentage polydispersities $\Delta$. ($\Delta$ measures  the width of a Gaussian distribution of particle radii.)  A selection of points from that data set is shown below in Fig.\,\ref{IGV2}.  As expected, the monodisperse system at $\Delta = 0\,\%$ exhibits an apparently sharp  transition between liquid and hexatic phases at $\phi \cong 0.69$.  With increasing $\Delta$, the transition points on the $p(\phi)$ curves move to larger $p$'s and $\phi$'s, and become less and less distinct.  They are invisible in the pressure data above $\Delta = 9 \%$, which is the value of the polydispersity for which  Tanaka et al.~\cite{TANAKAetal-10} report a bond-orientational correlation length that extrapolates to infinity at $\phi \cong 0.787$. The important point for present purposes is that the sequence of pressure curves in Fig.~2 of \cite{KAWASAKI-TANAKA-11} appears to indicate a smooth crossover from a liquid-hexatic transition at $\Delta = 0\,\%$ to Ising-like critical points for $\Delta \ge 9\,\%$ -- a qualitative change of universality class.

\section{Statistical Thermodynamics}
\label{Thermo}

In preparation for developing a model of thermalized, hard-core, glass forming particles, we need to understand the statistical thermodynamics of such systems.  The following analysis is based on \cite{BLI-09} and is almost, but not exactly, identical to that presented in \cite{LIEOU-JSL-12}.

In the absence of interaction energies, the only extensive quantity available for describing a hard-core system is its volume $V$.  (Throughout this paper, the term ``volume'' means either three dimensional volume or two dimensional area.) $V$ must be a function of the entropy $S$ plus a small set of internal state variables that govern the responses of the system to external forces.  Denote these variables by $\Lambda_{\alpha},~~\alpha = 1,...~n$ or, equivalently, by the set $\{\Lambda\}$.  As discussed in \cite{BLI-09}, the $\Lambda_{\alpha}$ must be extensive quantities, or spatial averages of such quantities, each carrying its own entropy.  In the thermodynamic limit of very large systems, the total entropy, say $S(V,\{\Lambda\})$, must approach its equilibrium value when the $\Lambda_{\alpha}$ approach their own equilibrium values. I discuss the choice of these variables in Section~\ref{Ising}; but, first, I consider only the general structure of the theory.  

For simplicity, assume that the kinetic degrees of freedom of the particles plus the degrees of freedom of the heat bath in which the system is immersed constitute a single thermal reservoir at temperature $\theta = k_B T$.  Denote the energy of this reservoir by $U_R$.  Because the hard-core configurational degrees of freedom carry no potential energy, $U_R$ is the total energy of the system.  Therefore, the first law of thermodynamics is simply $\dot U_R = -\,p\,\dot V$.  

The total entropy of this system, $S(V,\{\Lambda\})$, is the sum of the configurational entropy, say $S_C$, and the entropy of the reservoir, say $S_R$.  $S_C(V,\{\Lambda\})$ is a constrained entropy computed by counting the number of configurations with fixed values of $V$ and $\{\Lambda\}$.  Conversely, $V = V(S_C,\{\Lambda\})$.  In analogy to the notation of Edwards and coworkers  \cite{EDWARDS-89,MEHTA-EDWARDS-89,MEHTA-07}, define the compactivity $X$ by writing
\begin{equation}
X = \left({\partial V\over \partial S_C}\right)_{\{\Lambda\}}.
\end{equation}
Thus, the first law becomes
\begin{eqnarray}
\label{firstlaw}
\nonumber
\dot U_R &=& \theta\,\dot S_R= -\,p\,\dot V \cr\\&=& -\,p\,X\,\dot S_C -\,p\,\sum_{\alpha = 1}^n {\partial V\over \partial \Lambda_{\alpha}}\,\dot \Lambda_{\alpha}.
\end{eqnarray}
The second law, $\dot S_C + \dot S_R \ge 0$, is best written by using Eq.(\ref{firstlaw}) to eliminate $\dot S_C$, with the result
\begin{equation}
\label{secondlaw}
\dot S_R\,\left(1 - {\theta\over p\,X}\right) -\,{1\over X}\,\sum_{\alpha = 1}^n{\partial V\over \partial \Lambda_{\alpha}}\,\dot \Lambda_{\alpha} \,\ge 0.
\end{equation}
As usual, the requirement that this inequality be satisfied for arbitrary variations of external conditions implies that each of the terms on the right-hand side of Eq.(\ref{secondlaw}) be separately non-negative.  In particular, the first term is non-negative if
\begin{equation}
\dot S_R \propto 1 - {\theta\over p\,X},
\end{equation}
which means that, in the equilibrium limit, $X\to \theta/p$.   

To interpret the remaining terms on the right-hand side of Eq.(\ref{secondlaw}), remember that, in a statistical sense, the internal variables $\{\Lambda\}$ describe only a fraction of the number of configurational degrees of freedom of the system.  To account for the remaining internal degrees of freedom, write
\begin{equation}
\label{Stotal}
S_C(V,\{\Lambda\}) = {\cal S}(\{\Lambda\}) + S_1,
\end{equation}
and
\begin{eqnarray}
\label{Vtotal}
\nonumber
V(S_C,\{\Lambda\}) &=& {\cal V}(\{\Lambda\}) + V_1(S_1)\cr\\&=& {\cal V}(\{\Lambda\}) + V_1\Big[S_C -{\cal S}(\{\Lambda\})\Big],
\end{eqnarray}
where ${\cal S}(\{\Lambda\})$ is the entropy associated with the internal state variables; ${\cal V}(\{\Lambda\})$ is the corresponding volume; and $S_1$ and $V_1$ are, respectively, the entropy and volume associated with all the degrees of freedom other than the $\{\Lambda\}$.  Then, for each $\alpha$, 
\begin{equation}
\nonumber
{\partial V\over \partial \Lambda_{\alpha}} = {\partial {\cal V}\over \partial \Lambda_{\alpha}}- X\,{\partial {\cal S}\over \partial \Lambda_{\alpha}} = {\partial \over \partial \Lambda_{\alpha}}\,{\cal F}(\{\Lambda\}),
\end{equation}
where
\begin{equation}
\label{calF}
{\cal F}(\{\Lambda\}) = {\cal V}(\{\Lambda\})- X\,{\cal S}(\{\Lambda\}).
\end{equation}
Inserting this result into Eq.(\ref{secondlaw}), satisfying the second-law inequality separately for each term in the sum over $\alpha$, and taking the equilibrium limit, we find that 
\begin{equation}
\dot \Lambda_{\alpha} \propto -{\partial\over \partial \Lambda_{\alpha}}\,{\cal F}(\{\Lambda\}) \to 0.
\end{equation}
Thus, ${\cal F}(\{\Lambda\})$ is a ``free volume,'' analogous to a free energy, whose minimum in the space of variables $\{\Lambda\}$ locates the equilibrium state of the system. 
The values of the $\{\Lambda\}$ are determined by the equations 
\begin{equation}
\label{p-Lambda}
{p\over \theta} = {\partial {\cal S}/\partial \Lambda_{\alpha}\over \partial {\cal V}/\partial \Lambda_{\alpha}}.
\end{equation}

Note that Eq.(\ref{p-Lambda}) is what we would have found had we simply maximized the entropy ${\cal S}$ for a fixed volume ${\cal V}$, and used $X = \theta/p$ as a Lagrange multiplier. The preceding derivation is more general in the sense that it includes the residual quantities $S_1$ and $V_1$, which will play roles in the following analysis.

\section{Two-State Systems}
\label{Twostate}

The first step in constructing a model of hard-core particles within the framework outlined in Sec.~\ref{Thermo} must be to choose the internal state variables $\Lambda_{\alpha}$.  In doing this, we must decide which of the degrees of freedom of the system as a whole are statistically relevant, and therefore should be included among the $\Lambda_{\alpha}$, and which can be included implicitly in the residual quantities $S_1$ and $V_1$ defined in Eqs.(\ref{Stotal}) and (\ref{Vtotal}).  

In some way, one or more of these variables must describe the bond-orientational order observed by Tanaka {\it et al.}.\cite{TANAKAetal-10}  Tanaka's principal innovation has been to look for spatial correlations, not between particle positions {\it per se}, but between the positions of particles in topologically similar environments. Specifically, in what I believe to be the most definitive of their two-dimensional simulations \cite{TANAKAetal-10}, Tanaka {\it et al.} measured the time-averaged, complex, hexatic order parameter $\bar\Psi_6$ as a function of position and packing  fraction $\phi$; and they computed the two-point correlation $\langle\bar\Psi_6(r)\,\bar\Psi^*_6(0)\rangle$ as a function of the separation $r$.  From the latter quantity, they computed the correlation length $\xi(\phi)$, and found that it scaled as described in Sec.~\ref{Tanaka} as a function of $t \equiv (\phi_c - \phi)/\phi_c$, where the critical packing fraction $\phi_c$ has replaced $T_c^{-1}$.  Similar results were obtained in three dimensions, where the relevant topological order parameter was found to be the degree of hexagonal-close-packed  (as opposed to icosahedral) order. (See also \cite{TANAKA-12}.) 

It might seem that a topological order parameter such as $\bar\Psi_6$ would necessarily be one of the $\Lambda_{\alpha}$. Indeed, $\bar\Psi_6$ is frequently used as an argument of a Landau free energy, from which equilibrium states of two-dimensional systems are determined by a variational procedure formally identical to Eqs.(\ref{calF}) and (\ref{p-Lambda}). (For example, see \cite{NELSON-02}.) However, we need to look more closely at such models before trying to use them to describe glass formation.

In the monodisperse limit, Tanaka's hard disks undoubtedly undergo the liquid-hexatic transition that has been studied intensely ever since the pioneering numerical simulations of Alder and Wainwright.\cite{ALDER-WAINWRIGHT-62}  The standard description of such transitions in microscopically uniform materials is the two-dimensional melting theory of Kosterlitz, Thouless, Halperin, Nelson and Young \cite{KOSTERLITZ-THOULESS-72,KOSTERLITZ-74,HALPERIN-NELSON-78,NELSON-HALPERIN-79,YOUNG-79}; but this ``KTHNY'' theory may not be what we need to describe the most important properties of polydisperse, glass-forming liquids. KTHNY describes the melting of an hexatically ordered phase as a process in which a dilute population of disclination pairs undergoes a thermally induced unbinding transition, thereby destroying long-range orientational order in a distinctly non-Ising manner. In contrast, a glass-forming liquid, well away from a KTHNY transition, cannot naturally be described by a population of disclinations.  Even if it were possible to do so in some formal way, we know that the KTHNY analysis fails when that population becomes too dense, as must happen in the liquid phase.  Thus, it should be more productive to construct a theory in which topological order emerges from within a liquidlike state, instead of, as in KTHNY, starting from a state with infinitely long range order and asking how it melts. 

Accordingly, I propose that the fluctuating liquid state of a glass-forming  material be visualized as one in which topologically ordered clusters of particles appear and disappear in a background of disordered, fluidlike particles. These ordered clusters may be favored by steric (or energetic) interactions; in the liquid phase, they are disfavored by the entropy of the system as a whole.  As the pressure is increased, they come closer together, and the steric forces make it favorable for them to be aligned with each other. Thus, topological order grows with pressure. A mathematical description of how this happens is presented in Sec.\ref{Ising}. 

My main hypothesis is that these clusters are statistically most likely to be two-state systems.  The glass literature contains many references to such systems. For example, in 1972, Anderson, Halperin and Varma ~\cite{AHV-72} based their theory of low-temperature anomalies in glasses on the hypothesis that ``in any glass system there should be a certain number of atoms (or groups of atoms) which can sit more or less equally well in two equilibrium positions.''  More recently, my colleagues and I have used a similar argument to justify the model of two-state, shear transformation zones (STZ's) that we have used in theories of amorphous plasticity.\cite{FL-11} Twofold symmetry is especially important for present purposes because it is the Ising symmetry, and thus is consistent with the observed Ising-like critical exponents.  Moreover, the following argument in favor of twofold symmetry is independent of the specific nature of the ordering, or even the dimensionality, and thus may lead to the kind of universality that seems to be emerging in the computational experiments.

To see how two-state systems might naturally occur in theories of disordered materials, note that a spatially varying order parameter such as $\bar\Psi_6$ should be defined as an average over some coarse graining length scale.  If that scale is too small, say, only one or two particle spacings, then $\bar\Psi_6$ will be large in some places and small in others; but changing the local orientation at an hexatic site, i.e. changing the phase of $\bar\Psi_6$, almost certainly increases the local volume (or energy), so that the system is rigid at particle-sized length scales.  At the other extreme, if the coarse-graining scale is very large, and if we are not too close to an orientational ordering transition, then many different orientations will be degenerate in the sense of having the same volumes or energies, but the averaged magnitude of $\bar\Psi_6$ will be too small to provide useful information.  

An equivalent way of looking at this situation is to note that the small clusters of particles that play a role in forming bond-orientational order are those that have the sterically preferred topology, and {\it also} have enough flexibility to reorient themselves in the presence of their neighbors.  The minimum such flexibility is a two-fold orientational degeneracy.  More flexible small clusters, with higher order degeneracies, are statistically much less probable than the two-fold clusters. As a result, the natural coarse-graining scale -- the one that provides the statistically most relevant information -- is the one for which the ordered clusters are twofold degenerate, simply because two is the smallest integer greater than one.  

The coarse-graining argument immediately points to a limitation of the theory.  As an Ising system approaches a critical point, the correlations become long ranged, and a renormalization-group analysis like that used by KTHNY requires that we coarse-grain on increasingly large length scales. For hard disks in two dimensions, we eventually restore circular (``xy'') symmetry \cite{KOSTERLITZ-74}, and cross over into a regime where the KTHNY analysis again becomes valid.  As a result, even for a polydisperse system, there must be a region near a critical ordering transition where the correlation length diverges according to the KTHNY prediction. This crossover region may be unobservably small as a function of packing fraction for large polydispersities; but I think it must be there in principle.  Conversely, the crossover may also occur for a monodisperse system, because the liquid phase is intrinsically disordered away from criticality. 

A fundamental question regarding the two-state hypothesis is whether it can be derived systematically from a well defined description of a many-body system.  I see no reason why such a derivation should not be possible.  Perhaps the many-body strategy presented recently by Yaida \cite{YAIDA-12}, which also concludes that glassy systems belong to the Ising universality class, is a step in this direction.  However, in the next several Sections, I simply take the two-state model literally, and examine how its predictions compare with the simulation data.  

\section{Ising-Like Model}
\label{Ising}

\subsection{Binary Clusters}

To describe the two-state picture mathematically, let $N_+$ and $N_-$ be extensive variables denoting the numbers of, say, ``binary clusters'' oriented in $+$ and $-$ directions with respect to some direction in space. Degeneracy requires that, when a cluster switches between $+$ and $-$ orientations, it continues to make the same contribution, say $v^*$, to the volume ${\cal V}$ introduced in Eq.(\ref{Vtotal}). Just as in the STZ theory, the actual orientations denoted by $\pm$ need not be specified initially.  In the STZ case, we usually interpret the $N_{\pm}$ to be the numbers of zones whose orientations are more nearly parallel or antiparallel to an applied stress. In contrast to the STZ's (or the disclinations), there is no reason why the population of binary clusters should be dilute. A large ordered region at high compression may consist almost entirely of aligned clusters, whose specific orientation may be the result of an accidental anisotropy or a spontaneously broken symmetry.  

The way in which orientational order is propagated between neighboring positions in this model is via an Ising-like interaction, in which neighboring clusters make smaller contributions to the volume if their orientations are aligned than if they are opposite to each other. Another way of thinking about this is that the neighboring clusters break each other's orientational degeneracies in such a way as to increase the probability of their alignment.  This steric effect, the analog of an Ising exchange coupling, is the reason why orientational order increases in response to increasing pressure. 

To make direct comparisons with the functions $p(\phi)$, we need one more internal variable to describe how the system as a whole expands and contracts in response to changing pressure, and how that behavior couples to the internal state of topological order.  For this purpose, it is useful to introduce a population of, say, $N_0$ ``voids'' occupying volumes $v_0$.  In order to play a role comparable the the $N_{\pm}$, $N_0$ must be a collective variable describing a property of groups of particles comparable in size to the binary clusters, and $v_0$ must be a volume associated with more than just one, particle-size void. With the extra degree of freedom described by $N_0$, the model can make a transition with increasing pressure from dilute, liquidlike states with large populations of voids, to dense ordered states in which the voids disappear.  Tanaka et al. show such voids in their Voronoi tiling patterns for two dimensional systems (see Fig. 2 in \cite{TANAKAetal-10}), and assert that these voids play a role in limiting the extent of hexatic correlations.  That happens here as well. 

\subsection{Volume}
   
In a first, mean-field statement of this model, the volume ${\cal V}$ defined in Eq.(\ref{Vtotal}) is
\begin{equation}
\label{V}
{\cal V} \cong N^*\,v^* + N_0\,v_0 - {J\over 2\,(N^* + N_0)}\,(N_+^2+N_-^2),
\end{equation}
where  $N^* =N_++N_-$ is the total number of binary clusters, and is therefore proportional to the extensive number of statistically relevant, orientational degrees of freedom associated with the partial volume ${\cal V}$ and the partial entropy ${\cal S}$. The pairwise interaction, proportional to the ``exchange coupling'' $J$, is approximated here by the sum of the squares of the densities of the $\pm$ clusters. 

To see the analogy between Eq.(\ref{V}) and an Ising system, define 
\begin{equation}
\label{m-eta-def}
m = {N_+ - N_-\over N^*},~~~ \eta = {N^*\over N^* + N_0}.
\end{equation}
The variable $m$ is analogous to a magnetization; here, it is the bond-orientational order parameter. $\eta$ is a measure of how close the system is to its maximum density; it vanishes in the dilute limit, $N_0 \to \infty$, and goes to unity at high density where the voids are squeezed out of the system. In terms of these variables, Eq.(\ref{V}) becomes
\begin{equation}
\label{calV}
{{\cal V}(m,\eta)\over N^*}=  v^*  +\left({1\over \eta} - 1\right)\,v_0 - {1\over 4}\,J\,\eta\,(1 + m^2).
\end{equation}
Note that the term proportional to $J$ contains a factor $\eta$, implying that ordering becomes weaker with increased numbers of voids.  Note also that ${\cal V}(m,\eta)$ is proportional to $N^*$.  Because the partial entropy ${\cal S}$ also must be proportional to $N^*$, the latter quantity cancels out of the formula for the pressure, and there is no need to include it among the relevant internal variables. As a result, we can assume that $\{\Lambda\}$ consists of just the two variables $m$ and $\eta$.  

Equation (\ref{calV}) provides a formula for the packing fraction
\begin{equation}
 \phi = N_{tot}\,{\langle v \rangle\over V},
\end{equation} 
where $N_{tot}$ is the fixed total number of particles in the system, and $\langle v \rangle$ is the average volume of a single particle.  If we measure all volumes, including $J$, in units such that 
\begin{equation}
\label{<v>}
N_{tot}\,\langle v \rangle = N^*,
\end{equation}
then we can write  
\begin{equation}
\label{phi}
{1\over \phi} = \tilde v  +\left({1\over \eta} - 1\right)\,v_0 - {1\over 4}\,J\,\eta\,(1 + m^2).
\end{equation}
where $\tilde v = v^* + V_1/N^*$, and $V_1$ is the residual volume defined in Eq.(\ref{Vtotal}).  This scaling implies that both $\tilde v$ and $v_0$ are dimensionless numbers of the order of unity.  The total number of particles contained in $V_1$ must be proportional to $N_{tot}$, i.e. $V_1 \sim N_{tot}\,\langle v \rangle$; and therefore, with the volume units defined in Eq.(\ref{<v>}), $\tilde v \sim V_1/N^* \sim 1$.  Similarly, the total volume associated with voids must scale like $N_0\,v_0 \sim N_{tot}\,\langle v \rangle$.  Since we have required $N_0$ to scale with $N^*$ via Eq.(\ref{m-eta-def}), we again may use Eq.(\ref{<v>}) to find that $v_0 \sim 1$.

\subsection{Entropy}

Much of the physics of this model is contained in the choice of the entropy ${\cal S}$, defined in Eq.(\ref{Stotal}).  Like ${\cal V}$, ${\cal S}$ must be proportional to the number of statistically relevant degrees of freedom, $N^*$. Assume that ${\cal S}$ can be written in the form
\begin{equation}
{\cal S}(m,\eta)\cong {\cal S}_1(m) + {\cal S}_2(\eta). 
\end{equation}
The two-state model implies that ${\cal S}_1$ is an Ising-like function:
\begin{eqnarray}
\nonumber
{{\cal S}_1(m)\over N^*}&=& \ln\,(2) - {1\over 2}\,(1+m)\,\ln\,(1+m)\cr\\&-& {1\over 2}\,(1-m)\,\ln\,(1-m).~~~~~~~~~
\end{eqnarray}

The choice of ${\cal S}_2$ is more interesting but slightly problematic. If we make a lattice-gas approximation in which the $N_0$ voids are distributed randomly over $N^* + N_0$ sites, we find
\begin{equation}
{{\cal S}_2(\eta)\over N^*} \approx -\,\ln\,(\eta) - \left({1\over \eta}-1\right)\,\ln\,(1-\eta),
\end{equation}
which has an ideal gas limit as $\eta \to 0$, but vanishes very weakly as $\eta \to 1$. On the other hand, van der Waals behavior, with $p\sim \partial {\cal S}/\partial \eta \sim (1-\eta)^{-1}$, would require that ${\cal S}_2 \sim \ln\,(1 - \eta)$, which is unphysical because it diverges as $\eta \to 1$. 

My proposed alternative is
\begin{equation}
\label{calS2}
{{\cal S}_2(\eta)\over N^*} = -\,\ln\,(\eta)+ {A\over 1-\epsilon}\,(1-\eta)^{1-\epsilon},
\end{equation}
where the parameter $\epsilon$ must be in the range $0 < \epsilon < 1$.  Along with the adjustable parameter $A$, $\epsilon$ tunes the strength of the density dependence between weak and strong limits.  For small $\epsilon$, ${\cal S}_2$ approximates lattice gas behavior; as $\epsilon \to 1$, it resembles van der Waals.  In any case, the choice of ${\cal S}_2$ for large $\eta$ must be regarded as a phenomenological strategy for data fitting in the high-density limit, which is beyond the scope of the present investigation.  We know that this model lacks the ingredients for describing the way in which the system becomes glassy or crystalline at high densities. This is not a first-principles theory of such behavior; but it is useful to work with an approximate theory in which the high-density limit can be described.

\subsection{Spatial Variations}

A more general formulation of this theory starts with a partition function expressed as a functional integral over spatially varying values of $m$ and $\eta$:
\begin{equation}
\label{calZ}
{\cal Z}(X) = \int \delta m \int \delta \eta\,\,e^{- {\cal F}(m,\eta)/X}.
\end{equation}
Relations such as Eq.(\ref{p-Lambda}) can be interpreted as mean-field results, obtained by making a saddle-point approximation in Eq.(\ref{calZ}). As usual, write ${\cal F}/N^* \equiv f(m,\eta)$, and add a square-gradient term in the bond-orientational order parameter $m$: 
\begin{equation}
\label{fdef}
{{\cal F}\over N^*} \to f(m,\eta) + {\xi_0^2\over 2}\,(\nabla m)^2.
\end{equation}
Just as in the magnetic Ising model, the square-gradient term has its origins in the pairwise interactions proportional to $J$ in Eq.(\ref{V}). 

The standard procedure \cite{GOLDENFELD-92} for dealing with critical systems of this kind is to start with the unrenormalized form of of $f(m,\eta)$, expand it in powers of $m$ to obtain a Landau approximation, and then perform a renormalization-group analysis.  Beyond $m^4$, the higher powers in the expansion become irrelevant; and it is easy to check that the fluctuations in $\eta$ are non-critical. Thus, we know that this procedure produces the correct Ising scaling exponents for any set of starting parameters such that $f(m,\eta)$ has a mean-field critical point. Of course, this procedure tells us nothing about the possibility that the Ising symmetry might cross over to something else at large length scales.

%%%%%%%%%%%%% FIGURE 1 %%%%%%%%%%%%
\begin{figure}[here]
\centering \epsfig{width=.5\textwidth,file=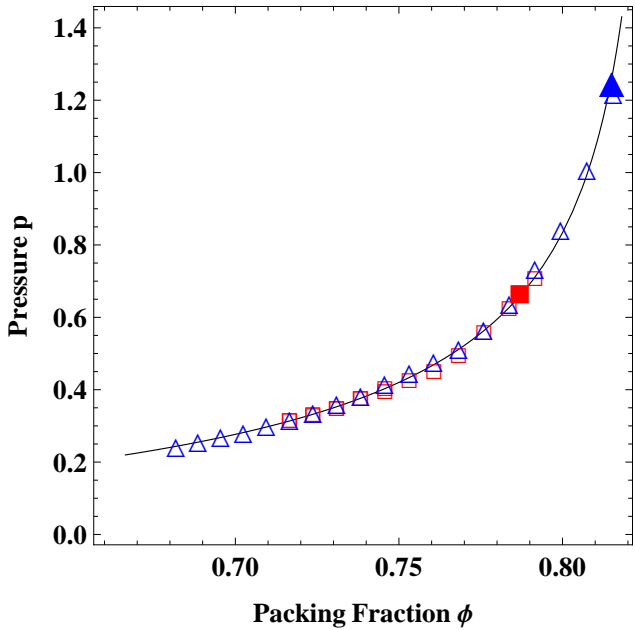} \caption{(Color online) Pressure as a function of packing fraction for polydispersties $\Delta = 9\,\%$ (red open squares) and $13\,\%$ (blue open triangles). The critical points are indicated by the corresponding solid symbols. The blue theoretical curve shows the $J= 0$ limit according to Eqs.(\ref{phiB}) and (\ref{pB}).  The pressure is expressed in dimensionless units defined in  \cite{KAWASAKI-TANAKA-11}. } \label{IGV1}
\end{figure}
%%%%%%%%%%%%%%%%%%%%%%%%%%%%%%%%%%%%% 

\section{Equations of State}
\label{Pphi}

We already know from \cite{TANAKAetal-10} and \cite{MOSAYEBIetal-10} that the correlation exponents for large enough polydispersities $\Delta$ are consistent with the renormalized Ising values $\nu \cong 2/d$.  Thus, the most interesting comparisons now are for the equations of state, $p(\phi)$, for different $\Delta$'s, reported in \cite{KAWASAKI-TANAKA-11}.  

The mean-field approximation using Eq.(\ref{p-Lambda}) means that we look for minima of $f(m,\eta)$ in the space of variables $m$ and $\eta$.  Thus, for variations with respect to $m$: 
\begin{equation}
\label{p-m}
{p\over \theta} = {1\over J\,\eta\,m}\,\ln\,\left({1+m\over 1-m}\right);
\end{equation}
and, for variations with respect to $\eta$:
\begin{equation}
\label{p-eta}
{p \over \theta} = {1\over v_0 + (J/4)\,\eta^2\,(1 + m^2)}\,\left[\eta + {A\,\eta^2\over (1- \eta)^\epsilon}\right].
\end{equation}
The reference volume $\tilde v$ defined in Eq.(\ref{phi}), and the parameters $v_0$, $A$ and $J$, should depend, at least in a first approximation, only on $\Delta$, and not on $\eta$. With known values of these parameters, Eqs.~(\ref{phi}), (\ref{p-m}), and (\ref{p-eta}) can be solved for $p/\theta$, $m$, and $\eta$ as functions of $\phi$.

Note that these equations recover a perfect gas law at very small values of $\eta$ and $m = 0$.  Eq.(\ref{p-eta}) implies that $p \approx \theta\,\eta/v_0$; and Eq.(\ref{phi}) implies that $\phi \approx \eta/v_0$.  Thus, as expected, $p \approx \theta\,\phi$. 

A second limiting behavior of these equations is especially interesting.  Large polydispersities $\Delta$ are roughly equivalent to high temperatures, which, in the magnetic analogy, imply small values of the coupling coefficient $J$.  Setting $J = 0$ in Eqs.(\ref{phi}) and (\ref{p-eta}), we find
\begin{equation}
\label{phiB}
{1\over \phi} \to \tilde v + \left({1\over \eta} - 1\right)\,v_0,
\end{equation}
and
\begin{equation}
\label{pB}
{p\,v_0\over \theta}\to  \eta + {A\,\eta^2\over (1-\eta)^{\epsilon}}.
\end{equation}
If the parameters $\tilde v$, $v_0$, $A$, and $\epsilon$ are independent of $\Delta$ in this limit, then the functions $p(\phi)$ should collapse onto a single curve.  I show in Fig.~\ref{IGV1} that this is nearly what happens for $\Delta = 9\,\%$ and $13\,\%$, the latter being the largest value for which I have data available.  This limiting behavior provides a convenient way to fix several of the theoretical parameters.   

The theoretical curve in Fig.~\ref{IGV1} is computed from Eqs.~(\ref{phiB}) and (\ref{pB}) as follows. I arbitrarily have set $v_0 = 1$, in accord with the argument following Eq.(\ref{phi}).  I also have set $\theta = 0.025$, which is the value given in  \cite{KAWASAKI-TANAKA-11}, where it is measured in units of the strength of the truncated, repulsive, Lennard-Jones potential used in those simulations, and therefore sets the scale for the pressure $p$. The remaining parameters have been chosen to fit the data. I find $\epsilon = 0.60$, about half way between the lattice gas and van der Waals limits, as discussed following Eq.(\ref{calS2}).  Like $v_0$ and $\theta$, I assume that $\epsilon$ is a constant, independent of $\Delta$.  According to Eq.(\ref{phiB}), the maximum packing fraction is $\phi_{max} = 1/\tilde v$.  For this large-$\Delta$ limit, and this choice of $\epsilon$, I find $\phi_{max} = 0.832$, and $A = 5.6$.

The Ising nature of Eqs.(\ref{phi}), (\ref{p-m}) and (\ref{p-eta}) appears in their reflection symmetry under $m \to -\,m$, which is spontaneously broken at critical points.  We can see this behavior at the mean-field level by expanding the logarithm in Eq.(\ref{p-m}) to third order in $m$ and solving the equation to find
\begin{equation}
\label{meanfieldm}
m \approx \cases {0, &for $\eta < \eta_c$ \cr \pm \sqrt{3(\eta/\eta_c-1)}, &for $\eta > \eta_c$,}
\end{equation}
where the critical value of $\eta$ is 
\begin{equation}
\label{etac}
\eta_c = {2\,\theta\over p_c\,J}, 
\end{equation}
and $p_c$ is the critical pressure.  This ``magnetization'' formula changes substantially under renormalization; the mean-field exponent implied by the square root, $\beta = 1/2$, changes to $\beta = 1/8$ for $d = 2$ and $\beta \cong 0.325$ for $d=3$.  This abrupt increase in $m$ for $\eta > \eta_c$ controls the behavior of the pressure at the onset of ordering {\it via} the quantity $m^2$ in the denominator of the right-hand side of Eq.(\ref{p-eta}).  Any meaningful comparison of Eq.(\ref{p-eta}) with the data requires that we use an expression for $m^2$ that is consistent with the renormalized theory.

To approximate the renormalized behavior, I have replaced the magnetization formula, Eq.(\ref{meanfieldm}), by one with the correct Ising exponent:
\begin{equation}
\label{Meta}
m \to M(\eta) = \cases{0, &for $\eta < \eta_c$ \cr \mu\,[(\eta/\eta_c-1)]^{\beta}, &for $\eta > \eta_c$\,,}
\end{equation}
and have used this approximation in Eqs.(\ref{phi}) and (\ref{p-eta}) instead of  the mean-field value of $m$ determined by Eq.({\ref{p-m}). In the absence of a simple, accurate interpolation from small to large values of $\eta/\eta_c-1$, I have used Eq.(\ref{Meta}) for all $\eta$, and have let $\mu$ be an adjustable parameter. The only vestige of Eq.({\ref{p-m}) in the theory is the formula for $\eta_c$ in Eq.(\ref{etac}), which is obtained from Eq.({\ref{p-m}) by taking the limit $m \to +0$.  We now may interpret $p_c$, $\phi_c$, and $\eta_c$ to be the renormalized values of those quantities; thus they are numbers that we can deduce directly from the data. 

%%%%%%%%%%%%% FIGURE 2 %%%%%%%%%%%%
\begin{figure}[here]
\centering \epsfig{width=.5\textwidth,file=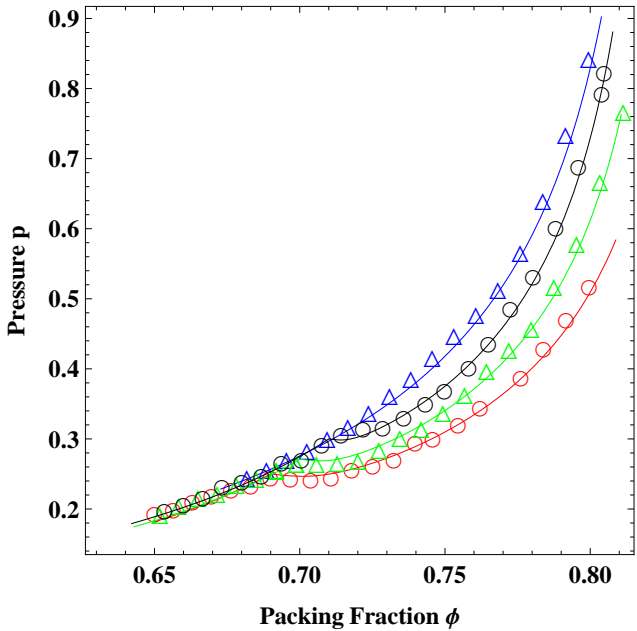} \caption{(Color online) Pressure as a function of packing fraction, from bottom to top, for polydispersties $\Delta = 0\,\%$ (red circles), $5\,\%$ (green triangles),  $~7\,\%$ (black circles), and $13\,\%$ (blue triangles). The corresponding solid curves indicate the theory described in the text. } \label{IGV2}
\end{figure}
%%%%%%%%%%%%%%%%%%%%%%%%%%%%%%%%%%%%% 

The results of these calculations, along with the simulation data from  \cite{KAWASAKI-TANAKA-11}, are shown in Fig.~\ref{IGV2} for selected polydispersities: $\Delta = 0\,\%, ~5\,\%,~7\,\%$, and $13\,\%$.  For $\Delta \le 8\,\%$, the values of $p_c$ and $\phi_c$ have been estimated  directly from the data, because the cusp-like changes in slope at the critical points are  visible.  For larger $\Delta$'s, $\phi_c$ has been evaluated in  \cite{KAWASAKI-TANAKA-11} by fitting relaxation-time measurements to a VFT formula; and the value of $\phi_c = 0.787$ for  $\Delta = 9\,\%$ also is consistent with the correlation-length measurements reported in \cite{TANAKAetal-10}.  The critical points for $\Delta = 9\,\%$ and $13\,\%$ are indicated respectively by the solid square and the solid triangle in Fig.~\ref{IGV1}. 

If we know $p_c$ and $\phi_c$ for a given $\Delta$, then Eqs.(\ref{phi}) and (\ref{p-eta}), evaluated at $m=0$ and $\eta = \eta_c$, plus the definition of $\eta_c$ in Eq.(\ref{etac}), provide three constraints on the five remaining unknown parameters: $\eta_c$, $\tilde v$, $J$, $A$, and $\mu$.   A fourth constraint is obtained by estimating $\phi_{max}$ by setting $\eta = 1$ in Eq.(\ref{phi}), but keeping nonzero $J$; that is
\begin{equation}
\label{phimax}
{1\over \phi_{max}} \cong \tilde v  - {J\over 4}\,\left[1 + M^2(1)\right].
\end{equation}
This is an approximate relation, because $M(\eta)$ given by Eq.(\ref{Meta}) is not accurate at $\eta = 1$; but Eq.(\ref{phimax}) is a useful consistency check on the earlier large-$\Delta$ estimate, especially in view of our lack of information about the glassy or ordered states in this limit. The only remaining free parameter in this analysis is $\mu$.  In computing the theoretical curves shown in Fig.\ref{IGV2}, I have used $\mu$ as the primary fitting parameter, and have kept $\phi_{max} = 0.832$ for all $\Delta$. 

%%%%%%%%%%%%% FIGURE 3 %%%%%%%%%%%%
\begin{figure}[here]
\centering \epsfig{width=.5\textwidth,file=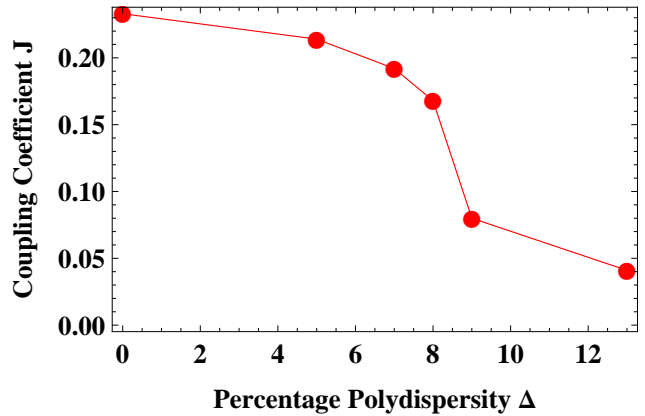} \caption{(Color online) Dimensionless coupling coefficient $J$ as a function of percentage polydispersity $\Delta$} \label{IGV3}
\end{figure}
%%%%%%%%%%%%%%%%%%%%%%%%%%%%%%%%%%%%

The comparisons between the theoretical curves and the data in Fig.\ref{IGV2}, and similar comparisons for other values of $\Delta$ not shown here, reveal physically plausible trends in the underlying parameters, especially the coupling coefficient $J$.  The main trend is that the critical points shift to higher values of $\phi$ with increasing $\Delta$, because increasing polydispersity suppresses the ordering transition.  The theoretical mechanism that produces this effect is the decrease in $J$ shown in Fig.~\ref{IGV3}.  In accord with the observations of Tanaka et al., $J$ drops abruptly at about $\Delta = 8\,\%$; but it does not drop to zero.  The values of $J$ for $\Delta = 9\,\%$ and $13\,\%$ have been computed as above, but using values for $\phi_c$ obtained from the correlation length and relaxation time analyses instead of directly from the $p(\phi)$ curves. This procedure produces values of $\tilde v$ and $A$ that are consistent with the $J = 0$ analysis, and does not visibly affect agreement with the data for $p(\phi)$; thus it serves as a consistency check.  The small but nonzero values of $J$ at large $\Delta$ produce Ising-like glass transitions. 

There are also more subtle effects.  The reference volume $\tilde v$ should be a slowly decreasing function of $\Delta$ because, at large polydispersities, smaller particles can fit into the spaces between larger particles in ways that do not happen in monodisperse systems.  The computed $\tilde v$ does decrease slightly from $1.33$ at  $\Delta = 0\,\%$ to $1.21$ at $\Delta = 13\,\%$.   The parameter $A$ is a measure of the entropy and, accordingly, increases with ``temperature'' $\Delta$ in this region, from $4.0$ to $5.6$, the latter value being the same as the one deduced from the $J = 0$ analysis.  The magnetization parameter $\mu$ decreases from $1.6$ at $\Delta = 0\,\%$ to about $1.0$ at $\Delta = 8\,\%$, but becomes impossible to measure at the larger $\Delta$'s where the critical transition is not visible in the $p(\phi)$ curves.  In plotting the curves for $\Delta = 9\,\%$ and $13\,\%$, I have simply set $\mu = 0$.

\section{Relaxation Rates}
\label{Rates}

In \cite{KAWASAKI-TANAKA-11}, Kawasaki and Tanaka devote much of their effort to measurements of the structural relaxation time $\tau_{\alpha}$, which they fit to a VFT function. They observe directly that $\log\,(\tau_{\alpha}) \sim \xi^{d/2}$, which tells us immediately that, while the glass transition may be Ising-like in its equilibrium behavior, it is qualitatively different dynamically. In comparison, the Hohenberg-Halperin \cite{HOHENBERG-HALPERIN-77} analysis tells us that the relaxation time for fluctuations of a non-conserved Ising magnetization diverges relatively weakly, like a power of $\xi$.  The difference is that relaxation events in a glass forming fluid near its transition point are highly collective phenomena, not amenable to the perturbation-theoretic methods or the assumptions about the nature of noise sources implicit in  \cite{HOHENBERG-HALPERIN-77} or in mode coupling theory. \cite{GOTZE91,GOTZE92}

For present purposes, assume that structural relaxations in glass forming materials occur at shear transformation zones (STZ's) \cite{FL-11,BL-11,JSL-12,JSL-EGAMI-12} or at other similarly soft locations.  The STZ's are naturally occurring structural defects that, in these hard-core systems, contain enough excess volume that they can undergo configurational rearrangements relatively easily.  If the characteristic excess volume of STZ's is $v_Z$, then their equilibrium population is proportional to a Boltzmann factor $\exp\,(-\,v_Z/X)$, where $X = \theta/p$.  To estimate a spontaneous STZ formation rate, and thus a relaxation rate, multiply this Boltzmann factor by an attempt frequency, $\rho(X)/\tau_0$, where $\tau_0$ is a microscopic time determined by the kinetic energies of the particles or the thermal fluctuations of the fluid in which they are suspended. Note that this picture of an activated process is already intrinsically nonperturbative.  The dimensionless attempt frequency $\rho(X)$ describes glassy slowing down as $X$ decreases. It is proportional to $\tau_0/\tau_{\alpha}$; and its evaluation is the goal of any glass theory.

Kawasaki and Tanaka  \cite{KAWASAKI-TANAKA-11} show by direct imaging that relaxation events occur primarily in disordered regions, consistent with the observation of Widmer-Cooper and Harrowell \cite{HARROWELL-07} that particles undergo rearrangements in regions of high ``propensity.'' In the present picture, this observation means simply that the STZ formation volume $v_Z$ is smaller in the disordered regions than in the ordered ones, so that the STZ's appear most frequently in the former.  However, the attempt frequency $\rho(X)$ must be a collective property of the system as a whole, rather than being determined just by the local environments of a few particles. 

The images of a glass forming fluid shown in  \cite{KAWASAKI-TANAKA-11} imply that correlated regions of size $\xi$ are slowly fluctuating into and out of existence, at a rate that I identify as being proportional to $\rho(X)/\tau_0$.  The STZ transitions provide the mechanism by which these fluctuations occur; conversely, it is these fluctuations that self-consistently generate the STZ's. To estimate this rate, note first that a correlated volume ${\cal V}_{corr}$ of linear size $\xi$ contains a number of particles proportional to $\xi^d$. In a thermally fluctuating system, each of these particles makes small, independent displacements through distances of the order of the interparticle spacing. Therefore,  ${\cal V}_{corr}$ undergoes Gaussian fluctuations of a characteristic magnitude $\delta\,{\cal V}_{corr}$ proportional to the square root of its size; that is, $\delta\,{\cal V}_{corr} \sim \xi^{d/2}$. To estimate a time scale for these fluctuations, note that they are slow, activated events.  Therefore, the statistical analysis in \cite{BLI-09,LIEOU-JSL-12} tells us that their frequency is proportional to
\begin{equation}
\label{rho-xi}
\rho(X) \sim  e^{-\,\delta{\cal V}_{c\!o\!r\!r}/X} \sim e^{-\,\xi^{d/2}/X_c} \sim e^{-\,1/t^w}, 
\end{equation}
where $w = d\,\nu/2= 1- \alpha/2 \cong 1$ for both $d= 2$ and $3$. Thus, if this rate is proportional to $\tau_{\alpha}^{-1}$, we recover the VFT formula.  In this way, I also learn that the XC approximation for computing $\rho(X)$ used too specialized a relaxation mechanism.

The Gaussian approximation made in deriving Eq.(\ref{rho-xi}) is similar to one made by Kirkpatrick et al.  in deriving the RFOT theory.\cite{RFOT-89}  Indeed, we may be discovering here why these two approaches to glass theory produce similar results.  Note, however, that the argument leading to Eq.(\ref{rho-xi}) assumes that we already know the diverging correlation length $\xi$, and then considers how that length determines the relaxation rate.  In RFOT, the Gaussian argument is used to determine the length scale itself on the basis of kinetic considerations. The term $\xi^{d/2}$ appears in RFOT as the effective surface energy of an entropically favored droplet. Moreover, Tanaka's picture of fluctuating regions of bond-orientational order seems qualitatively different from the mosaic structure postulated in RFOT and in its reinterpretation by Bouchaud and Biroli.\cite{BOUCHAUD-BIROLI-04}  

We can push the argument leading to Eq.(\ref{rho-xi}) a bit further by noting that it implies
\begin{equation}
\log\left({\tau_{\alpha}\over \tau_0}\right) \approx {D\,\phi\over \phi_c - \phi};~~~ D = p_c\,(a\,\xi_0)^{d/2}/\theta,
\end{equation} 
where $a$ is proportional to the particle spacing.  The bare correlation length $\xi_0$, defined in Eq.(\ref{fdef}), may be approximately the linear size of a binary cluster, and therefore ought to be a small multiple of $a$.  We know that $p_c$ increases with $\Delta$.  Thus, the ``fragility'' parameter $D$ is predicted to increase with $\Delta$ -- the glass becomes ``stronger'' -- in at least qualitative agreement with the increasing values of $D$ shown in the inset to Fig.~7 of \cite{KAWASAKI-TANAKA-11}.

\section{Concluding Remarks}  
\label{Conclusions}

My main hypothesis is that a population of statistically relevant, twofold-degenerate, ``binary clusters'' describes a broad class of  disordered systems in which the constituent particles have a tendency to develop some kind of local order.  If and when this hypothesis is correct, the system exhibits Ising-like behavior; in particular, correlation lengths associated with the favored ordering diverge with Ising-like critical exponents at glass transitions. These Ising correlations have been observed in several numerical simulations, primarily by Tanaka and coworkers.  As discussed in Sec.~\ref{Rates}, the observed Vogel-Fulcher-Tamann behavior of structural relaxation times also emerges from this two-state hypothesis. 

Tanaka's sequence of two-dimensional, hard-disk transitions, visible in the functions $p(\phi)$ shown here in Fig.~\ref{IGV2}, indicates a crossover from hexatic to glassy transitions with increasing polydispersity $\Delta$.  As discussed at the end of Sec.~\ref{Twostate}, however, we know that the theory fails in the close vicinity of these critical points, because the Ising symmetry must change to full circular symmetry when the correlations become sufficiently long ranged. We know from numerical studies by Jaster \cite{JASTER} that the monodisperse hard-disk system undergoes a KTHNY transition with a correlation length that grows exponentially near $\phi_c$, in contrast to the power-law growth characteristic of Ising systems.  We also know from Anderson et al. \cite{ANDERSONetal-12} that, when examined numerically with high precision extremely close to $\phi_c$, this transition is revealed to be weakly first order.  There is also an experimental study by Han et al. \cite{YODHetal-08}, who see KTHNY behavior for two-dimensional colloids at $3\,\%$ polydispersity. 

When looked at somewhat less closely, however, the two-state Ising theory appears to be remarkably successful.  It correctly predicts a sequence of critical ordering transitions with diverging correlation lengths, even for $\Delta = 0$.  The arguments in Sec.~\ref{Twostate}, if correct, make the Ising symmetry seem robust; there is no place in the two-state picture for a symmetry-breaking analog of a magnetic field.  The theory also makes roughly credible predictions for the ordered states at $\phi > \phi_c$, where the correlations again become short ranged, and mean-field approximations may regain validity.  The small, negative values of $dp/d\phi$ in the transition regions might be physically realistic indications of the weak, phase-separation instability reported in  \cite{ANDERSONetal-12}.  

The conjectured validity of mean-field approximations in the ordered regime might make it possible for some missing ingredients of the theory to be restored within the Ising-like formulation. For example, the theory in its present state contains no hint of translational order.  It does not tell us how or where to look for competition between glass formation and crystal growth. It resorts to a phenomenological expression for the entropy, in Eq.(\ref{calS2}), for computing the pressure at high packing fractions, where translational order should be present, at least for small $\Delta$.  It says nothing specific about the orientations of the local topologies, or the possibility of ``grain boundaries'' between topologically oriented regions.  It contains no information about how particles of different sizes are distributed spatially.  For example, in a bidisperse system of hard disks,  Donev et al. \cite{DONEVetal-06} found phase separation between the large and small particles in equilibrated structures at the highest densities.  If the two-state, Ising-like model does provide a reasonable starting approximation, then it might accomodate some of these other physical properties of glassy materials.

I emphasize, however, that this equilibrium theory of a glass forming liquid is definitely not a theory of the glassy state itself.  It might be a starting point for understanding how glass forming systems fall out of equilibrium and lose ergodicity during cooling or compression near their transition points; but it is not yet such a theory. \\

\begin{acknowledgments}

I thank H. Tanaka for providing the original simulation data, some of which is shown in Figs.~\ref{IGV1} and \ref{IGV2}, and most especially for thought provoking discussions about the issues addressed in this paper. This research  was supported in part by the U.S. Department of Energy, Office of Basic Energy Sciences, Materials Science and Engineering Division, DE-AC05-00OR-22725, through a subcontract from Oak Ridge National Laboratory.

\end{acknowledgments}

\end{document}